\pdfoutput=1
\documentclass[aps,prl,twocolumn,amsmath,amssym,superscriptaddress]{revtex4-1}
\usepackage{graphicx}
\usepackage{dcolumn}
\usepackage{bm}
\usepackage{url}
\usepackage{bm}   
\usepackage{bbm}   
\usepackage{verbatim}
\usepackage{stmaryrd}
\usepackage{amsthm}
\usepackage{amssymb}

\usepackage[usenames,dvipsnames]{color}
\newcommand{\sol}{}
\newcommand{\newsol}{} 
\newcommand{\unsol}{}
\newcommand{\unsolMinor}{}
\usepackage[normalem]{ulem}

\newcommand{\eq}{{\rm eq}}
\newcommand{\Neq}{{\rm neq}}
\newcommand{\add}{\unsol \rm add}

\newcommand{\Feq}{F}

\newcommand{\Fg}{F_{\unsol \Neq}}
\newcommand{\Wdiss}{W_{\rm diss}}
\newcommand{\Wex}{W_{\rm ex}}

\newcommand{\noneqFE}{{\unsol nonequilibrium free energy}}

\newcommand{\Ierase}{{\cal I}_{\rm \hspace{0.1ex} e}}

\newcommand{\letter}{paper}

\begin{document}\
\title{The thermodynamics of prediction}

\author{Susanne Still}
\affiliation{University of Hawaii at M\=anoa, Information and Computer Sciences, sstill@hawaii.edu}

\author{David A. Sivak}
\affiliation{Physical Biosciences Division, Lawrence Berkeley National Laboratory}

\author{Anthony J. Bell}
\affiliation{Redwood Center for Theoretical Neuroscience, University of California at Berkeley}

\author{Gavin E. Crooks}
\affiliation{Physical Biosciences Division, Lawrence Berkeley National Laboratory}

\begin{abstract}
A system responding to a stochastic driving signal 
can be interpreted as computing, by means of its dynamics, an 
implicit
model of the environmental variables. The system's state retains information about past environmental fluctuations, and a fraction of this information is predictive of future ones. The remaining nonpredictive information reflects model complexity that does not improve predictive power, 
and thus represents 
the ineffectiveness of the model. We expose the fundamental equivalence between this model inefficiency and thermodynamic inefficiency, measured by dissipation. Our results hold arbitrarily far from thermodynamic equilibrium and are applicable to a wide range of systems, including biomolecular machines. They highlight a profound connection between the effective use of information and efficient thermodynamic operation: any system constructed to keep memory about its environment {\em and} to operate 
with maximal energetic efficiency
{\em has to be predictive}.
\end{abstract}

\maketitle
All systems perform computations by means of responding to their environment. In particular, living systems compute, on a variety of length- and time-scales, future expectations based on their prior experience. 
Most
biological computation is fundamentally a nonequilibrium process, because a preponderance of biological machinery in its natural operation is driven far from thermodynamic equilibrium. For example, many molecular machines (such as the microtubule-associated motor kinesin) are driven by ATP hydrolysis, which liberates $\sim$500 meV per molecule~\cite{Lodish}. This energy is large compared with ambient thermal energy, $1~k_{\rm B}T\approx 25$~meV ($k_{\rm B}$ is Boltzmann's constant and the temperature is $T \sim$ 300~Kelvin). In general, such large energetic inputs drive the operative degrees of freedom of biological machines away from equilibrium averages.

Recently, significant progress has been made in describing driven systems far from equilibrium~\cite{Jarzynski:2011hl}, perhaps most notably Jarzynski's work relation \cite{Jarzy97} generalizing Clausius' Inequality, the further generalization embodied in fluctuation theorems \cite{Evans:1993tj,Crooks-fluct}, and the extension of these relations to calculating potentials of mean force~\cite{Hummer:2001hc}. These advances have allowed researchers to measure equilibrium quantities, such as free energy changes, by observing how a system reacts to being driven out of equilibrium, e.g.~\cite{Liphardt:2002cs,CollinJarBust2005}. 

This literature typically assumes that the experiment is known, i.e.\ that the exact time course of the driving signal is given. However, systems that are embedded in realistic environments, for example, a biological macromolecule operating under natural conditions, {\unsolMinor are exposed to stochastic driving}. Here, we therefore study driven systems for which the changes in the driving signal(s) are governed by some probability density $P_X$. 
{\sol This can be any stochastic process, and the results we derive require neither that $P_X$ has specific properties, nor that it is known by the system.} We assume that there is no feedback from the system to the driving signal. 
The dissipation, averaged not only over the system's path through its state space, but also over driving protocols, then quantifies the system's energetic inefficiency. 

The dynamics of the system perform a computation by changing the system's state, as a function of the driving signal. As a result, the new system state contains some memory about the driving signal. The system dynamics can be interpreted as computing a model: past environmental influences are mapped onto the current state of the system, which through its correlation with forthcoming environmental fluctuations implicitly contains a prediction of the future.

In this \letter, we ask how the quality of this (implicit) model is related to thermodynamic efficiency. But how do we measure the quality of a model? A useful model has to have predictive power (see e.g.~\cite{jeffreys, Bialek01, Still-IAL09, StillCruEl10}, and refs. therein), meaning it must capture predictive information \cite{shaw1984dripping, Crut88a, BNT01, CruFeld03}, while not being overly complicated. In other words, the model should contain as little dispensable {\em nonpredictive} information as possible.

Our central contribution is the demonstration of a fundamental equivalence between the instantaneous nonpredictive information carried by the system and the dissipation of energy.

\paragraph{Problem setup.--}
Let $s_t$ denote the state of the system at time~$t$, while $x_t$ denotes the driving signal. The dynamics of the system are modeled by discrete time Markovian conditional state-to-state transition probabilities, $p(s_t|s_{t-1},x_t)$. 
The external drive is governed by \mbox{$P_X=p(x_0,\dots,x_\tau)$}. We assume that at time $t=0$, the system is in thermodynamic equilibrium, in contact with a heat bath with inverse temperature $\beta:=1/k_{\rm B}T$. A change in the external driving signal $x_0\rightarrow x_1$ forces the system out of equilibrium. The system responds by changing its state $s_0\rightarrow s_1$, according to the transition probability $p(s_1|s_0,x_1)$. The external signal subsequently changes again $x_1\rightarrow x_2$, and the process is repeated until time $t=\tau$. 
\begin{center}
\includegraphics[]{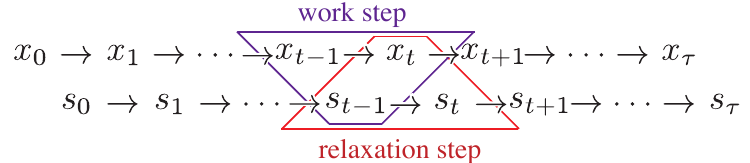}
\end{center}
The system remains in thermal contact with the heat bath during the entire protocol $x_0,\dots,x_\tau$, as in \cite{Crooks98}.
Work is done during a work step, as the external signal {\unsolMinor changes} from $x_{t-1}$ to $x_{t}$~\cite{Crooks98,Imparato}, 
\begin{equation}
W\![{\unsol s_{t-1};} x_{\unsol t-1} \! \rightarrow \! x_{\unsol t}] := E(s_{\unsol t-1},x_{\unsol t}) \!-\! E(s_{\unsol t-1},x_{\unsol t-1}) \ .
\label{work}
\end{equation}
In response to this change, the system relaxes from $s_{t-1}$ to $s_{t}$ in a relaxation step. The total work over the course of a driving protocol is 
\mbox{$W=\sum_{t=1}^{\tau}W[{\unsol s_{t-1};}x_{t-1}\!\rightarrow\!x_{t}]$}. 
The total change in energy, $\Delta E:=E(s_\tau,x_\tau)-E(s_0,x_0)$$=W+Q$, equals the total work plus the total heat, 
\mbox{$Q = \!\sum_{t=1}^{\tau} \!\left[E(s_{t},x_{t}) \!-\! E(s_{t-1},x_{t})\right]$}, 
flowing into the system during the relaxation steps. 

For now, we assume that the kernel which describes the dynamics, $p(s_t|s_{t-1},x_t)$, is fixed. However, the conditional distribution over states after the work step but before the system relaxes, $p(s_{t-1}|x_{t})$, changes as a function of time, as does the conditional distribution over states after the relaxation step, $p(s_{t}|x_{t})$. In general, these distributions are not the same, and neither of them is an equilibrium distribution. {\sol Under Markovian {\newsol system} dynamics, the probability before a relaxation step simplifies to
\begin{widetext}
\begin{align}
p(s_{t-1}|x_{t}) 
= \Bigg\langle \bigg\langle \Big\langle \cdots \big\langle p(s_{t-1}|s_{t-2}, x_{t-1}) \big\rangle_{p(s_{t-2}|s_{t-3}, x_{t-2})} 
\cdots \Big\rangle_{p(s_1|s_0, x_0)} \bigg\rangle_{p_{\rm eq}(s_0|x_0)} \Bigg\rangle_{p(x_0,\dots, x_{t-1}|x_t)} \ ,
\end{align}
\end{widetext}
and the distribution after relaxation is given by}
\begin{equation}
p(s_t|x_t) = \big\langle p(s_t| s_{t-1}, x_t) \big\rangle_{p(s_{t-1}|x_t)} \ .
\end{equation}
{\unsolMinor Angled brackets with a subscripted probability $p$ denote an average over $p$. 

{\unsolMinor The equilibrium distribution is the same function, before and after relaxation, $p_\eq(s|x_t):=e^{-\beta\left(E(s,x_t)-\Feq_t \right)}$, where $s$ refers to the state of the system with energy $E(s,x_t)$},
and $\Feq_t:=F[x_t]$ denotes equilibrium free energy. The probability of a specific path through the system's state space, given the protocol, is 
\begin{equation}
P_{S|X} = p_{\rm eq}(s_0|x_0) \prod_{t=1}^{\tau} p(s_t|s_{t-1}, x_t) \ , 
\end{equation} 
{\unsol and the joint probability} is
\begin{align}
P&_{S,X} := p(s_0, \dots, s_\tau, x_0, \dots, x_\tau) \\
&= p(x_0) p_{\rm eq}(s_0|x_0) \prod_{t=1}^{\tau} p(x_t|x_0,\dots,x_{t-1}) p(s_t|s_{t-1}, x_t) \ . \notag
\end{align}
In the following, unless otherwise clear from the context, angled brackets without a subscript denote an average over the distribution $P_{S,X}$. 

\paragraph{Dissipation out of equilibrium.--}
After the conclusion of the protocol, the probability over system states is given by {\unsolMinor $p(s_\tau|x_\tau)$,} in general not an equilibrium distribution. 
Then, in addition to the equilibrium free energy, $F_\tau$, there is {\unsolMinor another} contribution to the free energy, {\unsol because the system is not in thermodynamic equilibrium. This additional free energy} would be dissipated as heat to the environment if the system {\newsol were to relax} to thermodynamic equilibrium. {\newsol For Markovian system dynamics,} {\unsolMinor the} {\unsol additional} nonequilibrium contribution is proportional {\unsol \cite{shaw1984dripping}} to the relative entropy (Kullback-Leibler divergence) between the actual distribution 
{\unsolMinor $p(s_\tau|x_\tau)$}
at the end of the protocol and the equilibrium distribution, 
\begin{equation}
F_{\tau}^{\add}\big[ {\unsolMinor p(s_\tau|x_\tau)} \big] = k_B T \, D_{\rm KL}\big[ {\unsolMinor p(s_\tau|x_\tau)} \| p_{\rm eq}({\unsolMinor s_\tau}|x_\tau) \big] ~.
\end{equation}   
The non-negative Kullback-Leibler {\unsolMinor (KL)} divergence \cite{KL} between distributions $p(x)$ and $q(x)$ is defined as
\begin{equation}
D_{\rm KL}[ p(x) \| q(x) ] := \left\langle \ln \left[ \frac{p(x) }{ q(x)} \right] \right\rangle_{p(x)} \geq 0~.
\end{equation}
The sum of both contributions to the free energy {\unsolMinor constitutes} the {\unsol overall \em nonequilibrium} free energy, $\Fg\left[{\unsolMinor p(s_\tau|x_\tau)}\right]{\unsolMinor =}F_\tau+F_{\tau}^{\add}\left[ {\unsolMinor p(s_\tau|x_\tau)}\right]$.
{\unsolMinor Here,} \noneqFE\ {\unsolMinor is defined} in analogy to the standard equilibrium free energy as a functional of the probability distribution, but applied to any probability distribution \cite{GavSchul97,Crooks07,SivakCrooks}, {\unsolMinor that is} to any $p(s|x)$:
\begin{equation}
\Fg \left[p(s|x) \right] := \!{\unsolMinor \langle E(s,x) \rangle_{p(s|x)} \!+\! k_B T \langle \ln [p(s|x)] \rangle_{p(s|x)}}. \!\!\! \label{FGDef}
\end{equation}
{\unsol The average work {\em irretrievably} lost 
over the course of a
driving protocol, 
\begin{equation}
\langle \Wdiss \rangle_{P_{S|X}} := \langle W \rangle_{P_{S|X}} - \Delta \Fg  \label{WdissDef}~,
\end{equation}
equals the average work {\unsolMinor performed on the system} minus the} {\unsol nonequilibrium} free energy change $\Delta\Fg:=\Fg\left[{\unsolMinor p(s_\tau|x_\tau)}\right]-\Fg\left[{\unsolMinor p(s_0|x_0)}\right]$. We {\unsolMinor can compare this to the average}  {\em excess work} {\unsolMinor for a given protocol,}
{\unsol $\langle\Wex\rangle_{P_{S|X}}:=\langle W\rangle_{P_{S|X}}-\Delta\Feq$}, the total work done on the system in excess of the equilibrium free energy change $\Delta\Feq\!:=\!F_\tau\!-\!F_0$ which would be the work done if the driving signal changed quasistatically (infinitely slowly){\unsolMinor, and hence the system remained in thermodynamic equilibrium throughout the protocol}.  {\unsolMinor This excess work equals the dissipated work only when the protocol includes a final equilibration of the system.} 

Since the system starts in equilibrium, the total change in \noneqFE\ is the equilibrium free energy change plus the abovementioned {\unsol additional} contribution to the free energy, $\Delta\Fg=\Delta F+F_{\tau}^{\add}\left[{\unsolMinor p(s_\tau|x_\tau)}\right]$. The dissipation is then the excess work minus this {\unsol additional} nonequilibrium contribution,
\begin{equation}
\langle \Wdiss \rangle_{\!\unsol P_{S|X}} \!\!=\! \langle \Wex \rangle_{\!\unsol P_{S|X}} \!-\!
F_{\tau}^{\add}\!\left[{\unsolMinor p(s_\tau|x_\tau)}\right] 
\!\leq\! \langle \Wex \rangle_{\!\unsol P_{S|X}}.\!\!\! \label{wdissgleqwdiss}
\end{equation}
{\unsol Later, we derive a lower bound on the dissipation and excess work averaged over all protocols, denoted by $\langle\Wdiss\rangle$, and $\left\langle\Wex\right\rangle$, respectively.} 

Each of the incremental work steps, $x_t\!\rightarrow\!x_{t+1}$, is accompanied by a \noneqFE\ change given by 
$\Delta\Fg[x_t\!\rightarrow\!x_{t+1}]\!:={\unsolMinor\!\Fg[p(s_t|x_{t+1})]\!-\!\Fg[p(s_t|x_t)]}$, 
so that the average dissipation during each work step is
{\unsol
\begin{align}
\big\langle \Wdiss [x_t \! \rightarrow \! x_{t+1}] \big\rangle &:= \big\langle W [s_t;x_t \! \rightarrow \! x_{t+1}] \big\rangle_{p(s_t,x_t,x_{t+1})}\! \notag\\ 
&\,\,- \big\langle\Delta \Fg [x_t \! \rightarrow \! x_{t+1}] \big\rangle_{p(x_t,x_{t+1})}.  \label{diss-t}
\end{align}
The \noneqFE\ change during each relaxation step is 
{\unsol 
\begin{align}
\!\!\! \Delta \Fg [x_t;s_{t-1} \! \rightarrow \! s_{t}] \!=\! F_{t}^{\add}[p(s_{t}|x_{t})] \!-\! F_{t}^{\add}[p(s_{t-1}|x_{t})],\!
\end{align}
which 
equals
the change in KL divergence from the equilibrium distribution.}

\paragraph{Predictive power, memory, and dissipation.--}
The system state and the external signal are random variables that potentially share information,
{\unsol $I[s_t,x_t]:=\left\langle\ln\left[p(s_t,x_t)/p(s_t)p(x_t)\right]\right\rangle_{p(s_t, x_t)}$, where $p(s_t)=\langle p(s_t|x_t)\rangle_{p(x_t)}$. {\unsol Mutual information \cite{shannon48} measures the reduction in uncertainty about the outcome of a random variable upon learning the identity of another variable, and it is symmetric: $I[s_t,x_t]=H[s_t]-H[s_t|x_t]=H[x_t]-H[x_t|s_t]$. Uncertainty is quantified by the entropy, $H[s_{t}]:=-\langle\ln p(s_{t})\rangle_{p(s_{t})}$, and the conditional entropy, $H[s_{t}|x_t]:=-\langle\ln p(s_{t}|x_t)\rangle_{p(s_{t},x_t)}$, respectively.}}
 
{\sol The system transition probability, $p(s_t|s_{t-1},x_t)$, is assumed to depend on the current signal value $x_t$ and system state $s_{t-1}$. These two dependencies are sufficient to induce correlations between the system's current state and previous signal values. The memory the system keeps about the external signal can then be quantified by the information that the system state 
$s_t$
retains about a trajectory $\{x_{t-\tau_m},\dots,x_t\}$. In general, there are temporal correlations in the input signal, and hence, there can be correlations between 
$s_t$
and future signal values. That is, some of the memory retained in the system's state} is information about the future trajectory $\{x_{t+1},\dots,x_{t+\tau_f}\}$. 
{\unsolMinor Here we focus on the {\em instantaneous} memory, $I_{\rm mem}(t):=I[s_t,x_t]$, and the {\em instantaneous} predictive power \cite{Still-IAL09}, }
{\unsolMinor $I_{\rm pred}(t):=I[s_t,x_{t+1}]=H[x_{t+1}]-H[x_{t+1}|s_t]$.}

The implicit model computed by the system's dynamics which map signal $x_t$ onto state $s_t$, given the current state $s_{t-1}$, contains the probabilistic map $p(x_{t+1}|s_t)$,
which represents the prediction of $x_{t+1}$, given the value of $s_t$. 

The {\unsolMinor instantaneous {\em nonpredictive information} is} defined as the difference between instantaneous memory and predictive power, $I_{{\rm mem}}(t)-I_{{\rm pred}}(t)$. {\unsolMinor It} represents useless nostalgia and provides a measure for the ineffectiveness of the model.  

Averaging the \noneqFE\ over protocols allows us to write 
\begin{equation}
\beta \big\langle \Fg[p(s|x)] \big\rangle_{p(x)} = \beta \big\langle E(s,x) \big\rangle_{p(s,x)} - {\unsolMinor H[s|x]} \ . \label{FGave}
\end{equation}
{\unsolMinor Combining this with Eqs.~(\ref{work}) and (\ref{diss-t})}
\footnote{{\sol $\beta\left<\Wdiss[x_t \shortrightarrow x_{t+1}] \right> \\= \beta \left( \big\langle E(s_t,x_{t+1}) \big\rangle_{p(s_t,x_{t+1})} - \big\langle E(s_t,x_t) \big\rangle_{p(s_t,x_t)}\right) \\- \beta \left(\big\langle \Fg[p(s|x_{t+1})] \big\rangle_{p(x_{t+1})} + \big\langle \Fg[p(s|x_t)] \big\rangle_{p(x_t)} \right) \\= H[s_t|x_{t+1}] - H[s_t|x_{t}] = I_{{\rm mem}}(t) - I_{{\rm pred}}(t)$.}}, 
we arrive at our first result: the {\unsol instantaneous} nonpredictive information is proportional to the average work dissipated while the signal changes from $x_t$ to $x_{t+1}$,
\begin{equation}
\beta \big\langle \Wdiss[x_t  \shortrightarrow x_{t+1}] \big\rangle= I_{{\rm mem}}(t) - I_{{\rm pred}}(t) \ . \label{InfLoss} 
\end{equation}
In summary, the unwarranted retention of past information is fundamentally equivalent to energetic inefficiency.

\paragraph{Lower bound on total dissipation.--}
We now relate the total average dissipated work during the entire protocol, {\unsol averaged over all protocols, $\langle\Wdiss\rangle$}, to the total nostalgia, $I_{\rm mem}-I_{\rm pred}$, given by the difference between the total instantaneous memory, \mbox{$I_{\rm mem}:=\sum_{t=0}^{\tau-1}I_{{\rm mem}}(t)$}, and the total instantaneous predictive power, \mbox{$I_{\rm pred}:=\sum_{t=0}^{\tau-1}I_{{\rm pred}}(t)$}. 
{\unsol To that end we 
need to combine Eqs.~\eqref{diss-t} and \eqref{InfLoss}, and 
sum over all time steps. This sum includes a sum over changes in \noneqFE, which can be expressed as}
\begin{equation}
\left\langle \sum_{t=0}^{\tau - 1} \Delta \Fg[x_t \! \rightarrow \! x_{t+1}] \right\rangle = {\unsolMinor \big\langle} \Delta \Fg  - \Delta F^{\rm relax}_{\unsol \rm neq} \big\rangle
\end{equation}
in terms of the total \noneqFE\ change, ${\unsol\langle\Delta\Fg\rangle}$, and the sum of \noneqFE\ changes during relaxation steps:
\begin{equation}
\big\langle \Delta F^{\rm relax}_{\unsol \rm neq} \big\rangle :=  \left\langle \sum_{t=0}^{\tau - 1} \Delta \Fg[x_t;s_t \rightarrow s_{t+1}] \right\rangle \leq 0 ~. \label{FR}
\end{equation}
This quantity is nonpositive because, on average, during relaxation steps the system evolves toward equilibrium. 
The total dissipation then becomes, using Eq.~(\ref{InfLoss}),
\begin{equation}
\beta \langle \Wdiss \rangle = I_{\rm mem} - I_{\rm pred} - \beta \big\langle \Delta F^{\rm relax}_{\unsol \rm neq}  \big\rangle . \label{result-compact}
\end{equation}
The total nostalgia therefore provides a lower bound on the total average dissipation, and also, due to Eq.~(\ref{wdissgleqwdiss}), on the total average excess work, 
\begin{eqnarray}
I_{\rm mem} - I_{\rm pred} \leq \beta \langle  \Wdiss \rangle \leq  \beta \langle  \Wex \rangle~. \label{inequality}
\end{eqnarray}

We can use this result to refine Landauer's principle~\cite{Landauer1961}, which states that any erasure of information must be balanced by an increase in entropy elsewhere. The information erased during a protocol, such as the reset protocol of Landauer~\cite{Landauer1961}, is the entropy change $\Ierase:=H[s_0|x_0]-H[s_\tau|x_\tau]$. Note that the information erased here is not mutual information about the driving signal, but rather information that could have potentially been extracted from the system by some measurement process. Landauer pointed out that the erasure of information requires heat to flow out of the system, which obeys (using the first and second laws of thermodynamics, and Eqs.~(\ref{FGDef}) and (\ref{WdissDef})) 
\begin{align}
- \beta \langle Q \rangle &= \Ierase + \beta \langle \Wdiss \rangle \geq \Ierase ~.\label{deltaS} 
\end{align}
Substituting our result from Eq.~\eqref{result-compact} into Eq.~\eqref{deltaS} yields the new relation
\begin{align}
- \beta \langle Q \rangle &= \Ierase + \ I_{\rm mem} - I_{\rm pred} - \beta \big\langle \Delta F^{\rm relax}_{\unsol \rm neq} \big\rangle \ . 
\end{align}
Thus (using Eq.~\eqref{FR}) we obtain a refinement of Landauer's principle,
\begin{equation}
- \beta \langle Q \rangle \geq \Ierase  + I_{\rm mem} - I_{\rm pred} \ ,
\end{equation}
where the bound is augmented by the total nostalgia. The system dynamics of a computing device that retains memory therefore must be maximally predictive to approach Landauer's limit.

\paragraph{Discussion.--}
{\sol The dynamics, $p(s_t|s_{t-1},x_t)$, have been assumed fixed for any given system. However, biological systems are typically adapted to their environment. One can then ask if there is a simple principle underlying the process producing this adaptation. If such a principle exists, then it may reflect the importance of energetic efficiency, because of the resulting competitive advantage for reproducing organisms. While other criteria play a role, such as robustness and sensitivity, energetic efficiency is of fundamental 
{\newsol relevance}. 
This is exemplified by biological molecules that harness environmental fluctuations to accomplish energetically-costly downstream tasks. The more efficiently such a molecule can operate, the more it can accomplish  through coupling to endergonic chemical reactions or mechanical actions. For example, with more efficient coupling to the environment, the molecular motor kinesin can carry larger cargos. Likewise, with greater efficiency cytochrome c oxidase complex, an enzyme that pumps protons across a membrane, can create stronger electrochemical gradients. Evidence for the importance of energetic efficiency is furthermore found in biomolecular machines that approach 100\% efficiency when driven in a natural fashion: the stall torque for the F$_1$-ATPase~\cite{Kinosita:2000} and the stall force for Myosin V~\cite{Cappello:2007} are near the maximal values possible given the free energy liberated by ATP hydrolysis and the sizes of their respective rotations and steps.

These and many other biological functions require some correspondence between the environment and the systems that implement them. Therefore the memory of their instantiating systems must be nonzero. We have shown that any such system with nonzero memory 
must conduct predictive inference, at least implicitly, to approach maximal energetic efficiency. 

A substantial amount of work has sought to relate emerging biological functions and behaviors to efficient energy usage. Examples range from animal behavior (e.g. \cite{Alexander_82L,*Alexander_81}) to single neurons, where recently researchers have proposed that the minimization of energy expenditure, subject to constraints given by the desired function, may be ``a unifying principle governing neuronal biophysics''~\cite{sejn}. On the other hand, there is much research on 
 optimal information processing in neurons. For a recent review see e.g. \cite{Bialek12}, which proposes that the extraction of predictive information in biological signal processing may constitute, or at least lead to, a general principle.
By directly relating memory and predictive power to dissipation out of equilibrium, the results we have derived here indicate that these two important paradigms are deeply connected. 

While it is perhaps intuitive that neurons and organisms should have to implement predictive inference to function well, our results have the striking implication that on all scales energetic efficiency calls for predictive inference. This includes the smallest biological units, such as molecular machines.

Our results also specify the required kind of predictive inference: maximization of predictive power at a desired level of system memory, as in \cite{Still-IAL09}. This connects with work on optimal predictive inference algorithms discussed in \cite{Crut88a,Still-IAL09,StillCruEl10,StillPrecup11}, and references therein.
We envision implementing these algorithms in fast and efficient hardware. The results we have derived here could then be used to choose the energetically most efficient implementation among the many possible choices. }

\paragraph{Conclusion.--} 

\newsol{
We argued that dissipation far from thermodynamic equilibrium {\unsolMinor is given by average} work minus {\em {\unsol nonequilibrium}} free energy change. We also argued that the {\unsol nonpredictive part of a system's memory provides a natural measure for the inefficiency of a system's implicit model of its environment.}

We showed that instantaneous nonpredictive information
is proportional to the energy dissipated when an external driving signal changes by an incremental amount, thereby doing work on the system. This result demonstrates the intimate connection between prediction and energetic efficiency. Summed over the entire protocol, the total nonpredictive information
provides a lower bound on the total dissipation. 

{\unsolMinor These} results imply that any system which is built to have nonzero memory {\em has to be predictive} in order to allow for minimal possible dissipation, i.e.\ {\em to operate {\unsol at maximal energetic efficiency.}} 
{\unsol 
Our results furthermore 
allow for
a refinement of Landauer's principle, applied to systems away from thermodynamic equilibrium.}

We have provided} a connection between nonequilibrium thermodynamics and learning theory, {\unsol by making precise how two important aspects of life are fundamentally related: making a predictive model of the environment and using available energy efficiently.}
}

\paragraph{Acknowledgements.--}
SS thanks William Bialek and Robert S. Shaw for many inspiring discussions, and Matteo Marsili for helpful technical comments. SS and AJB are thankful for conversations with James P. Crutchfield and Michael R. DeWeese. DAS and GEC were funded by the Office of Basic Energy Sciences of the U.S. Department of Energy under Contract No. DE-AC02-05CH11231. AJB was funded by NSF grants SBE 0542013 and SMA 1041755 to the Temporal Dynamics of Learning Center, an NSF Science of Learning Center.


\begin{thebibliography}{32}%
\makeatletter
\providecommand \@ifxundefined [1]{%
 \@ifx{#1\undefined}
}%
\providecommand \@ifnum [1]{%
 \ifnum #1\expandafter \@firstoftwo
 \else \expandafter \@secondoftwo
 \fi
}%
\providecommand \@ifx [1]{%
 \ifx #1\expandafter \@firstoftwo
 \else \expandafter \@secondoftwo
 \fi
}%
\providecommand \natexlab [1]{#1}%
\providecommand \enquote  [1]{``#1''}%
\providecommand \bibnamefont  [1]{#1}%
\providecommand \bibfnamefont [1]{#1}%
\providecommand \citenamefont [1]{#1}%
\providecommand \href@noop [0]{\@secondoftwo}%
\providecommand \href [0]{\begingroup \@sanitize@url \@href}%
\providecommand \@href[1]{\@@startlink{#1}\@@href}%
\providecommand \@@href[1]{\endgroup#1\@@endlink}%
\providecommand \@sanitize@url [0]{\catcode `\\12\catcode `\$12\catcode
  `\&12\catcode `\#12\catcode `\^12\catcode `\_12\catcode `\%12\relax}%
\providecommand \@@startlink[1]{}%
\providecommand \@@endlink[0]{}%
\providecommand \url  [0]{\begingroup\@sanitize@url \@url }%
\providecommand \@url [1]{\endgroup\@href {#1}{\urlprefix }}%
\providecommand \urlprefix  [0]{URL }%
\providecommand \Eprint [0]{\href }%
\@ifxundefined \urlstyle {%
  \providecommand \doi  [0]{\begingroup \@sanitize@url \@doi}%
  \providecommand \@doi [1]{\endgroup \@@startlink {\doibase
  #1}doi:\discretionary {}{}{}#1\@@endlink }%
}{%
  \providecommand \doi  [0]{doi:\discretionary{}{}{}\begingroup
  \urlstyle{rm}\Url }%
}%
\providecommand \doibase [0]{http://dx.doi.org/}%
\providecommand \Doi [0]{\begingroup \@sanitize@url \@Doi }%
\providecommand \@Doi  [1]{\endgroup\@@startlink{\doibase#1}\@@Doi}%
\providecommand \@@Doi [1]{#1\@@endlink}%
\providecommand \selectlanguage [0]{\@gobble}%
\providecommand \bibinfo  [0]{\@secondoftwo}%
\providecommand \bibfield  [0]{\@secondoftwo}%
\providecommand \translation [1]{[#1]}%
\providecommand \BibitemOpen [0]{}%
\providecommand \bibitemStop [0]{}%
\providecommand \bibitemNoStop [0]{.\EOS\space}%
\providecommand \EOS [0]{\spacefactor3000\relax}%
\providecommand \BibitemShut  [1]{\csname bibitem#1\endcsname}%
\bibitem [{\citenamefont {Lodish}\ \emph {et~al.}(2000)\citenamefont {Lodish},
  \citenamefont {Berk}, \citenamefont {Zipursky}, \citenamefont {Matsudaira},
  \citenamefont {Baltimore},\ and\ \citenamefont {Darnell}}]{Lodish}%
  \BibitemOpen
  \bibfield  {author} {\bibinfo {author} {\bibfnamefont {H.}~\bibnamefont
  {Lodish}}, \bibinfo {author} {\bibfnamefont {A.}~\bibnamefont {Berk}},
  \bibinfo {author} {\bibfnamefont {S.}~\bibnamefont {Zipursky}}, \bibinfo
  {author} {\bibfnamefont {P.}~\bibnamefont {Matsudaira}}, \bibinfo {author}
  {\bibfnamefont {D.}~\bibnamefont {Baltimore}}, \ and\ \bibinfo {author}
  {\bibfnamefont {J.}~\bibnamefont {Darnell}},\ }\href@noop {} {\emph {\bibinfo
  {title} {Molecular Cell Biology}}}\ (\bibinfo  {publisher} {W.H. Freeman, San Francisco},\
  \bibinfo {year} {2000})\BibitemShut {NoStop}%
\bibitem [{\citenamefont {Jarzynski}(2011)}]{Jarzynski:2011hl}%
  \BibitemOpen
  \bibfield  {author} {\bibinfo {author} {\bibfnamefont {C.}~\bibnamefont
  {Jarzynski}},\ }\href@noop {} {\bibfield  {journal} {\bibinfo  {journal}
  {Annu. Rev. Condens. Matter Phys.},\ }\textbf {\bibinfo {volume} {2}},\
  \bibinfo {pages} {329} (\bibinfo {year} {2011})}\BibitemShut {NoStop}%
\bibitem [{\citenamefont {Jarzynski}(1997)}]{Jarzy97}%
  \BibitemOpen
  \bibfield  {author} {\bibinfo {author} {\bibfnamefont {C.}~\bibnamefont
  {Jarzynski}},\ }\href@noop {} {\bibfield  {journal} {\bibinfo  {journal}
  {Phys. Rev. Lett.},\ }\textbf {\bibinfo {volume} {78}},\ \bibinfo {pages}
  {2690} (\bibinfo {year} {1997})}\BibitemShut {NoStop}%
\bibitem [{\citenamefont {Evans}\ \emph {et~al.}(1993)\citenamefont {Evans},
  \citenamefont {Cohen},\ and\ \citenamefont {Morriss}}]{Evans:1993tj}%
  \BibitemOpen
  \bibfield  {author} {\bibinfo {author} {\bibfnamefont {D.~J.}\ \bibnamefont
  {Evans}}, \bibinfo {author} {\bibfnamefont {E.~G.~D.}\ \bibnamefont {Cohen}},
  \ and\ \bibinfo {author} {\bibfnamefont {G.~P.}\ \bibnamefont {Morriss}},\
  }\href@noop {} {\bibfield  {journal} {\bibinfo  {journal} {Phys. Rev.
  Lett.},\ }\textbf {\bibinfo {volume} {71}},\ \bibinfo {pages} {2401}
  (\bibinfo {year} {1993})}\BibitemShut {NoStop}%
\bibitem [{\citenamefont {Crooks}(1999)}]{Crooks-fluct}%
  \BibitemOpen
  \bibfield  {author} {\bibinfo {author} {\bibfnamefont {G.~E.}\ \bibnamefont
  {Crooks}},\ }\href@noop {} {\bibfield  {journal} {\bibinfo  {journal} {Phys.
  Rev. E},\ }\textbf {\bibinfo {volume} {60}},\ \bibinfo {pages} {2721}
  (\bibinfo {year} {1999})}\BibitemShut {NoStop}%
\bibitem [{\citenamefont {Hummer}\ and\ \citenamefont
  {Szabo}(2001)}]{Hummer:2001hc}%
  \BibitemOpen
  \bibfield  {author} {\bibinfo {author} {\bibfnamefont {G.}~\bibnamefont
  {Hummer}}\ and\ \bibinfo {author} {\bibfnamefont {A.}~\bibnamefont {Szabo}},\
  }\href@noop {} {\bibfield  {journal} {\bibinfo  {journal} {P. Natl. Acad.
  Sci. USA},\ }\textbf {\bibinfo {volume} {98}},\ \bibinfo {pages} {3658}
  (\bibinfo {year} {2001})}\BibitemShut {NoStop}%
\bibitem [{\citenamefont {Liphardt}\ \emph {et~al.}(2002)\citenamefont
  {Liphardt}, \citenamefont {Dumont}, \citenamefont {Smith}, \citenamefont
  {Tinoco},\ and\ \citenamefont {Bustamante}}]{Liphardt:2002cs}%
  \BibitemOpen
  \bibfield  {author} {\bibinfo {author} {\bibfnamefont {J.~T.}\ \bibnamefont
  {Liphardt}}, \bibinfo {author} {\bibfnamefont {S.}~\bibnamefont {Dumont}},
  \bibinfo {author} {\bibfnamefont {S.~B.}\ \bibnamefont {Smith}}, \bibinfo
  {author} {\bibfnamefont {I.}~\bibnamefont {Tinoco}}, \ and\ \bibinfo {author}
  {\bibfnamefont {C.}~\bibnamefont {Bustamante}},\ }\href@noop {} {\bibfield
  {journal} {\bibinfo  {journal} {Science},\ }\textbf {\bibinfo {volume}
  {296}},\ \bibinfo {pages} {1832} (\bibinfo {year} {2002})}\BibitemShut
  {NoStop}%
\bibitem [{\citenamefont {Collin}\ \emph {et~al.}(2005)\citenamefont {Collin},
  \citenamefont {Ritort}, \citenamefont {Jarzynski}, \citenamefont {Smith},
  \citenamefont {Tinoco},\ and\ \citenamefont
  {Bustamante}}]{CollinJarBust2005}%
  \BibitemOpen
  \bibfield  {author} {\bibinfo {author} {\bibfnamefont {D.}~\bibnamefont
  {Collin}}, \bibinfo {author} {\bibfnamefont {F.}~\bibnamefont {Ritort}},
  \bibinfo {author} {\bibfnamefont {C.}~\bibnamefont {Jarzynski}}, \bibinfo
  {author} {\bibfnamefont {S.~B.}\ \bibnamefont {Smith}}, \bibinfo {author}
  {\bibfnamefont {I.}~\bibnamefont {Tinoco}}, \ and\ \bibinfo {author}
  {\bibfnamefont {C.}~\bibnamefont {Bustamante}},\ }\href@noop {} {\bibfield
  {journal} {\bibinfo  {journal} {Nature},\ }\textbf {\bibinfo {volume}
  {437}},\ \bibinfo {pages} {231} (\bibinfo {year} {2005})}\BibitemShut
  {NoStop}%
\bibitem [{\citenamefont {Jeffreys}(1998)}]{jeffreys}%
  \BibitemOpen
  \bibfield  {author} {\bibinfo {author} {\bibfnamefont {H.}~\bibnamefont
  {Jeffreys}},\ }\href@noop {} {\emph {\bibinfo {title} {{Theory of
  Probability}}}},\ \bibinfo {edition} {3rd}\ ed.\ (\bibinfo  {publisher}
  {Oxford University Press},\ \bibinfo {year} {1998})\BibitemShut {NoStop}%
\bibitem [{\citenamefont {Bialek}(2001)}]{Bialek01}%
  \BibitemOpen
  \bibfield  {author} {\bibinfo {author} {\bibfnamefont {W.}~\bibnamefont
  {Bialek}},\ }in\ \href@noop {} {\emph {\bibinfo {booktitle} {Physics of
  bio-molecules and cells, Proceedings of the Les Houches Summer 
  School, Session LXXV}}}\ (\bibinfo  {publisher} {Springer Verlag},\ \bibinfo {year}
  {2001})\ pp.\ \bibinfo {pages} {485--577}\BibitemShut {NoStop}%
\bibitem [{\citenamefont {Still}(2009)}]{Still-IAL09}%
  \BibitemOpen
  \bibfield  {author} {\bibinfo {author} {\bibfnamefont {S.}~\bibnamefont
  {Still}},\ }\href@noop {} {\bibfield  {journal} {\bibinfo  {journal}
  {EPL},\ }\textbf {\bibinfo {volume} {85}},\ \bibinfo {pages}
  {28005} (\bibinfo {year} {2009})}\BibitemShut {NoStop}%
\bibitem [{\citenamefont {Still}\ \emph {et~al.}(2010)\citenamefont {Still},
  \citenamefont {Crutchfield},\ and\ \citenamefont {Ellison}}]{StillCruEl10}%
  \BibitemOpen
  \bibfield  {author} {\bibinfo {author} {\bibfnamefont {S.}~\bibnamefont
  {Still}}, \bibinfo {author} {\bibfnamefont {J.~P.}\ \bibnamefont
  {Crutchfield}}, \ and\ \bibinfo {author} {\bibfnamefont {C.}~\bibnamefont
  {Ellison}},\ }\href@noop {} {\bibfield  {journal} {\bibinfo  {journal}
  {Chaos},\ }\textbf {\bibinfo {volume} {20}},\ \bibinfo {pages} {037111}
  (\bibinfo {year} {2010})}\BibitemShut {NoStop}%
\bibitem [{\citenamefont {Shaw}(1984)}]{shaw1984dripping}%
  \BibitemOpen
  \bibfield  {author} {\bibinfo {author} {\bibfnamefont {R.}~\bibnamefont
  {Shaw}},\ }\href@noop {} {\emph {\bibinfo {title} {{The dripping faucet as a
  model chaotic system}}}}\ (\bibinfo  {publisher} {Aerial Press, Santa Cruz},\ \bibinfo
  {year} {1984})\BibitemShut {NoStop}%
\bibitem [{\citenamefont {Crutchfield}\ and\ \citenamefont
  {Young}(1989)}]{Crut88a}%
  \BibitemOpen
  \bibfield  {author} {\bibinfo {author} {\bibfnamefont {J.~P.}\ \bibnamefont
  {Crutchfield}}\ and\ \bibinfo {author} {\bibfnamefont {K.}~\bibnamefont
  {Young}},\ }\href@noop {} {\bibfield  {journal} {\bibinfo  {journal} {Phys.
  Rev. Lett.},\ }\textbf {\bibinfo {volume} {63}},\ \bibinfo {pages} {105}
  (\bibinfo {year} {1989})}\BibitemShut {NoStop}%
\bibitem [{\citenamefont {Bialek}\ \emph {et~al.}(2001)\citenamefont {Bialek},
  \citenamefont {Nemenman},\ and\ \citenamefont {Tishby}}]{BNT01}%
  \BibitemOpen
  \bibfield  {author} {\bibinfo {author} {\bibfnamefont {W.}~\bibnamefont
  {Bialek}}, \bibinfo {author} {\bibfnamefont {I.}~\bibnamefont {Nemenman}}, \
  and\ \bibinfo {author} {\bibfnamefont {N.}~\bibnamefont {Tishby}},\
  }\href@noop {} {\bibfield  {journal} {\bibinfo  {journal} {Neural Comput.},\
  }\textbf {\bibinfo {volume} {13}},\ \bibinfo {pages} {2409} (\bibinfo {year}
  {2001})}\BibitemShut {NoStop}%
\bibitem [{\citenamefont {Crutchfield}\ and\ \citenamefont
  {Feldman}(2003)}]{CruFeld03}%
  \BibitemOpen
  \bibfield  {author} {\bibinfo {author} {\bibfnamefont {J.~P.}\ \bibnamefont
  {Crutchfield}}\ and\ \bibinfo {author} {\bibfnamefont {D.~P.}\ \bibnamefont
  {Feldman}},\ }\href@noop {} {\bibfield  {journal} {\bibinfo  {journal}
  {Chaos},\ }\textbf {\bibinfo {volume} {13}},\ \bibinfo {pages} {25 }
  (\bibinfo {year} {2003})}\BibitemShut {NoStop}%
\bibitem [{\citenamefont {Crooks}(1998)}]{Crooks98}%
  \BibitemOpen
  \bibfield  {author} {\bibinfo {author} {\bibfnamefont {G.~E.}\ \bibnamefont
  {Crooks}},\ }\href@noop {} {\bibfield  {journal} {\bibinfo  {journal} {J.
  Stat. Phys.},\ }\textbf {\bibinfo {volume} {90}},\ \bibinfo {pages} {1481}
  (\bibinfo {year} {1998})}\BibitemShut {NoStop}%
\bibitem [{\citenamefont {Imparato}\ \emph {et~al.}(2007)\citenamefont
  {Imparato}, \citenamefont {Peliti}, \citenamefont {Pesce}, \citenamefont
  {Rusciano},\ and\ \citenamefont {Sasso}}]{Imparato}%
  \BibitemOpen
  \bibfield  {author} {\bibinfo {author} {\bibfnamefont {A.}~\bibnamefont
  {Imparato}}, \bibinfo {author} {\bibfnamefont {L.}~\bibnamefont {Peliti}},
  \bibinfo {author} {\bibfnamefont {G.}~\bibnamefont {Pesce}}, \bibinfo
  {author} {\bibfnamefont {G.}~\bibnamefont {Rusciano}}, \ and\ \bibinfo
  {author} {\bibfnamefont {A.}~\bibnamefont {Sasso}},\ }\href@noop {}
  {\bibfield  {journal} {\bibinfo  {journal} {Phys. Rev. E},\ }\textbf
  {\bibinfo {volume} {76}},\ \bibinfo {pages} {050101} (\bibinfo {year}
  {2007})}\BibitemShut {NoStop}%
\bibitem [{\citenamefont {Kullback and Leibler}(1948)}]{KL}%
  \BibitemOpen
  \bibfield  {author} {\bibinfo {author} {\bibfnamefont {S.}\ \bibnamefont
  {Kullback}}\ and\ \bibinfo {author} {\bibfnamefont {R.~A.}\ \bibnamefont
  {Leibler}},\ }\href@noop {} {\bibfield  {journal} {\bibinfo  {journal} {Ann. 
  Math. Stat.},\ }\textbf {\bibinfo {volume} {22}},\ \bibinfo {pages} {79}
  (\bibinfo {year} {1951})}\BibitemShut {NoStop}  
\bibitem [{\citenamefont {Gaveau}\ and\ \citenamefont
  {Schulman}(1997)}]{GavSchul97}%
  \BibitemOpen
  \bibfield  {author} {\bibinfo {author} {\bibfnamefont {B.}~\bibnamefont
  {Gaveau}}\ and\ \bibinfo {author} {\bibfnamefont {L.~S.}\ \bibnamefont
  {Schulman}},\ }\href@noop {} {\bibfield  {journal} {\bibinfo  {journal}
  {Phys. Lett. A},\ }\textbf {\bibinfo {volume} {229}},\ \bibinfo {pages} {347}
  (\bibinfo {year} {1997})}\BibitemShut {NoStop}%
\bibitem [{\citenamefont {Crooks}(2007)}]{Crooks07}%
  \BibitemOpen
  \bibfield  {author} {\bibinfo {author} {\bibfnamefont {G.~E.}\ \bibnamefont
  {Crooks}},\ }\href@noop {} {\bibfield  {journal} {\bibinfo  {journal} {Phys.
  Rev. E},\ }\textbf {\bibinfo {volume} {75}},\ \bibinfo {pages} {041119}
  (\bibinfo {year} {2007})}\BibitemShut {NoStop}%
\bibitem [{\citenamefont {Sivak}\ and\ \citenamefont
  {Crooks}(2012)}]{SivakCrooks}%
  \BibitemOpen
  \bibfield  {author} {\bibinfo {author} {\bibfnamefont {D.~A.}\ \bibnamefont
  {Sivak}}\ and\ \bibinfo {author} {\bibfnamefont {G.~E.}\ \bibnamefont
  {Crooks}},\ }\href@noop {} {\bibfield  {journal} {\bibinfo  {journal} {Phys.
  Rev. Lett.},\ }\textbf {\bibinfo {volume} {108}},\ \bibinfo {pages} {150601}
  (\bibinfo {year} {2012})}\BibitemShut {NoStop}%
\bibitem [{\citenamefont {Shannon}(1948)}]{shannon48}%
  \BibitemOpen
  \bibfield  {author} {\bibinfo {author} {\bibfnamefont {C.~E.}\ \bibnamefont
  {Shannon}},\ }\href@noop {} {\bibfield  {journal} {\bibinfo  {journal} {Bell
  Syst. Tech. J.},\ }\textbf {\bibinfo {volume} {27}},\ \bibinfo {pages} {379}
  (\bibinfo {year} {1948})}\BibitemShut {NoStop}%
 \bibitem [{Note1()}]{Note1}%
  \BibitemOpen
  \bibinfo {note} {{\protect \color {black}$\beta \left <W_{\protect \rm
  diss}[x_t \shortrightarrow x_{t+1}] \right > \\= \beta \left ( {\setbox \z@
  \hbox {\frozen@everymath \@emptytoks \mathsurround \z@ $\nulldelimiterspace
  \z@ \left \delimiter "426830A \vcenter to\@ne \big@size {}\right .$}\box \z@
  }E(s_t,x_{t+1}) {\setbox \z@ \hbox {\frozen@everymath \@emptytoks
  \mathsurround \z@ $\nulldelimiterspace \z@ \left \delimiter "526930B \vcenter
  to\@ne \big@size {}\right .$}\box \z@ }_{p(s_t,x_{t+1})} - {\setbox \z@ \hbox
  {\frozen@everymath \@emptytoks \mathsurround \z@ $\nulldelimiterspace \z@
  \left \delimiter "426830A \vcenter to\@ne \big@size {}\right .$}\box \z@
  }E(s_t,x_t) {\setbox \z@ \hbox {\frozen@everymath \@emptytoks \mathsurround
  \z@ $\nulldelimiterspace \z@ \left \delimiter "526930B \vcenter to\@ne
  \big@size {}\right .$}\box \z@ }_{p(s_t,x_t)}\right ) \\- \beta \left
  ({\setbox \z@ \hbox {\frozen@everymath \@emptytoks \mathsurround \z@
  $\nulldelimiterspace \z@ \left \delimiter "426830A \vcenter to\@ne \big@size
  {}\right .$}\box \z@ }F_{{\protect \rm neq}}[p(s_t|x_{t+1})] {\setbox \z@ \hbox
  {\frozen@everymath \@emptytoks \mathsurround \z@ $\nulldelimiterspace \z@
  \left \delimiter "526930B \vcenter to\@ne \big@size {}\right .$}\box \z@
  }_{p(x_{t+1})} - {\setbox \z@ \hbox {\frozen@everymath \@emptytoks
  \mathsurround \z@ $\nulldelimiterspace \z@ \left \delimiter "426830A \vcenter
  to\@ne \big@size {}\right .$}\box \z@ }F_{{\protect \rm neq}}[p(s_t|x_t)]
  {\setbox \z@ \hbox {\frozen@everymath \@emptytoks \mathsurround \z@
  $\nulldelimiterspace \z@ \left \delimiter "526930B \vcenter to\@ne \big@size
  {}\right .$}\box \z@ }_{p(x_t)} \right ) \\= H[s_t|x_{t+1}] - H[s_t|x_{t}] = 
  I_{{\protect \rm mem}}(t) - I_{{\protect \rm pred}}(t)$.}}\BibitemShut
  {Stop}%
\bibitem [{\citenamefont {Landauer}(1961)}]{Landauer1961}%
  \BibitemOpen
  \bibfield  {author} {\bibinfo {author} {\bibfnamefont {R.}~\bibnamefont
  {Landauer}},\ }\href@noop {} {\bibfield  {journal} {\bibinfo  {journal} {IBM
  J. Res. Develop.},\ }\textbf {\bibinfo {volume} {5}},\ \bibinfo {pages} {183}
  (\bibinfo {year} {1961})}\BibitemShut {NoStop}%
\bibitem [{\citenamefont {Kinosita}\ \emph {et~al.}(2000)\citenamefont
  {Kinosita}, \citenamefont {Yasuda}, \citenamefont {Noji},\ and\ \citenamefont
  {Adachi}}]{Kinosita:2000}%
  \BibitemOpen
  \bibfield  {author} {\bibinfo {author} {\bibfnamefont {K.}~\bibnamefont
  {Kinosita}}, \bibinfo {author} {\bibfnamefont {R.}~\bibnamefont {Yasuda}},
  \bibinfo {author} {\bibfnamefont {H.}~\bibnamefont {Noji}}, \ and\ \bibinfo
  {author} {\bibfnamefont {K.}~\bibnamefont {Adachi}},\ }\href@noop {}
  {\bibfield  {journal} {\bibinfo  {journal} {Philos. T. Roy. Soc. B},\
  }\textbf {\bibinfo {volume} {355}},\ \bibinfo {pages} {473} (\bibinfo {year}
  {2000})}\BibitemShut {NoStop}%
\bibitem [{\citenamefont {Cappello}\ \emph {et~al.}(2007)\citenamefont
  {Cappello}, \citenamefont {Pierobon}, \citenamefont {Symonds}, \citenamefont
  {Busoni}, \citenamefont {Gebhardt}, \citenamefont {Rief},\ and\ \citenamefont
  {Prost}}]{Cappello:2007}%
  \BibitemOpen
  \bibfield  {author} {\bibinfo {author} {\bibfnamefont {G.}~\bibnamefont
  {Cappello}}, \bibinfo {author} {\bibfnamefont {P.}~\bibnamefont {Pierobon}},
  \bibinfo {author} {\bibfnamefont {C.}~\bibnamefont {Symonds}}, \bibinfo
  {author} {\bibfnamefont {L.}~\bibnamefont {Busoni}}, \bibinfo {author}
  {\bibfnamefont {J.~C.~M.}\ \bibnamefont {Gebhardt}}, \bibinfo {author}
  {\bibfnamefont {M.}~\bibnamefont {Rief}}, \ and\ \bibinfo {author}
  {\bibfnamefont {J.}~\bibnamefont {Prost}},\ }\href@noop {} {\bibfield
  {journal} {\bibinfo  {journal} {P. Natl. Acad. Sci. USA},\ }\textbf {\bibinfo
  {volume} {104}},\ \bibinfo {pages} {15328} (\bibinfo {year}
  {2007})}\BibitemShut {NoStop}%
\bibitem [{\citenamefont {Alexander}(1982)}]{Alexander_82L}%
  \BibitemOpen
  \bibfield  {author} {\bibinfo {author} {\bibfnamefont {R.~M.}\ \bibnamefont
  {Alexander}},\ }\href@noop {} {\emph {\bibinfo {title} {{Locomotion of
  Animals}}}}\ (\bibinfo  {publisher} {Blackie, Glasgow and London},\ \bibinfo
  {year} {1982})\BibitemShut {NoStop}%
\bibitem [{\citenamefont {Alexander}(1981)}]{Alexander_81}%
  \BibitemOpen
  \bibfield  {author} {\bibinfo {author} {\bibfnamefont {R.~M.}\ \bibnamefont
  {Alexander}},\ }\href@noop {} {\emph {\bibinfo {title} {{Optima for
  Animals}}}}\ (\bibinfo  {publisher} {Edward Arnold, London},\ \bibinfo {year}
  {1981})\BibitemShut {NoStop}%
\bibitem [{\citenamefont {Hasenstaub}\ \emph {et~al.}(2010)\citenamefont
  {Hasenstaub}, \citenamefont {Otte}, \citenamefont {Callaway},\ and\
  \citenamefont {Sejnowski}}]{sejn}%
  \BibitemOpen
  \bibfield  {author} {\bibinfo {author} {\bibfnamefont {A.}~\bibnamefont
  {Hasenstaub}}, \bibinfo {author} {\bibfnamefont {S.}~\bibnamefont {Otte}},
  \bibinfo {author} {\bibfnamefont {E.}~\bibnamefont {Callaway}}, \ and\
  \bibinfo {author} {\bibfnamefont {T.~J.}\ \bibnamefont {Sejnowski}},\
  }\href@noop {} {\bibfield  {journal} {\bibinfo  {journal} {P. Natl. Acad.
  Sci. USA},\ }\textbf {\bibinfo {volume} {107}},\ \bibinfo {pages} {12329}
  (\bibinfo {year} {2010})}\BibitemShut {NoStop}%
\bibitem [{\citenamefont {Bialek}()}]{Bialek12}%
  \BibitemOpen
  \bibfield  {author} {\bibinfo {author} {\bibfnamefont {W.}~\bibnamefont
  {Bialek}},\ }\href@noop {} {\enquote {\bibinfo {title} {Biophysics: Searching
  for principles},}\ }\bibinfo {note} {(Princeton University, Princeton, NJ, to be published)}
  \BibitemShut {NoStop}%
\bibitem [{\citenamefont {Still}\ and\ \citenamefont
  {Precup}(2011)}]{StillPrecup11}%
  \BibitemOpen
  \bibfield  {author} {\bibinfo {author} {\bibfnamefont {S.}~\bibnamefont
  {Still}}\ and\ \bibinfo {author} {\bibfnamefont {D.}~\bibnamefont {Precup}},\
  }\href@noop {} {\bibfield  {journal} {\bibinfo  {journal} {GSO 2009 Proc.; in Theory Biosci.}} 
  \textbf {\bibinfo {volume} {131}},\ \bibinfo {pages} {139} (\bibinfo {year} {2012})}\
  \bibinfo {note} {}
  \BibitemShut
  {NoStop}%
\end{thebibliography}
%

\end{document}